# DUCT: An Interactive Define-Use Chain Navigation Tool for Relative Debugging

Aaron Searle †                John Gough[†]                David Abramson[§]

*† Centre for Information Technology Innovation*
*Queensland University of Technology*
*Brisbane, Queensland, Australia*

*{a.searle,j.gough}@qut.edu.au*

*§ School of Compute Science & Software Engineering*
*Monash University*
*Clayton, Victoria, Australia*

davida@csse.monash.edu.au

## Abstract

*This paper describes an interactive tool that facilitates following define-use chains in large codes. The motivation for the work is to support "relative debugging", where it is necessary to iteratively refine a set of assertions between different versions of a program. DUCT is novel because it exploits the Microsoft Intermediate Language (MSIL) that underpins the .NET Framework. Accordingly, it works on a wide range of programming languages without any modification. The paper describes the design and implementation of DUCT, and then illustrates its use with a small case study.*

## 1. Introduction

In 1994 Abramson and Sosic invented a new debugging paradigm called "Relative Debugging" [1,2,3], and developed a reference implementation called Guard [23]. Relative debugging allows a user to both test and debug an application against a working reference version. It supports the idea of software evolution in which a program is changed incrementally, possibly because of changes in the program function, but also because of external changes in the environment, such as different computer architectures, operating system and run time libraries, and languages. Relative debugging facilitates the comparison, at run time, of the contents of key data structures. The methodology supports iteratively refining a program code until the source of a divergence can be discovered. This approach means that even large programs can be reduced quickly to small regions of code that behave differently. A number of case studies have illustrated the power of the approach [1,2].

The most flexible method for comparing data structure contents in Guard is implemented by user specified, declarative *assertions*. These commands state that a particular data structure in one program, at one line, should be the same as another data structure in another program at a different line. The methodology for deciding where to place assertions is built around following the data and control flow of the codes. For example, if the output of a computation is observed to be incorrect, then the user typically refines the assertions to follow the inputs of that computation – typically by finding the definition points of the variables in the right hand side of the computation. Traditionally, Guard has provided no support for this tracing, and has simply relied on the user being able to traverse the code to find these definition points of interest.

The early versions of Guard were command line driven, and thus, required the user to maintain separate windows onto the two source programs and the debugger. Recently, Guard has been integrated into a number of different interactive development environments (IDEs), namely Microsoft's Visual Studio [20], IBM's Eclipse [10] and Sun's Sun One Studio [24]. All of these offer significant advantages to Guard as they allow the user to maintain two separate source code views concurrently, and even side-by-side in the environment. Further, specifying assertions is easier, because it is possible to point and click to the variables and lines of interest rather than needing to find the line numbers and manually enter this information into an assertion command. The prototype version of Guard under Microsoft's Visual Studio is called VSGuard.

Having integrated Guard into an IDE we were interested in adding support to assist with identifying and navigating the definition points of the data structures, rather than requiring the user to perform this operation manually. This paper introduces a tool, called DUCT, which provides such support by identifying the definitions that reach a particular variable use within a Microsoft .NET program.

DUCT is novel because it only performs interactive analysis on specified variables, making it fast. Further, program analysis is performed directly on the .NET program which allows DUCT to function with any programming language that targets the Microsoft .NET framework. Finally, embedding DUCT in an IDE allows a user to navigate complex programs quickly, making relative debugging significantly easier than in the past. DUCT also allows users to trace data definitions whilst executing the program under debugger control, providing a unique and powerful platform for locating errors.

The paper is organized as follows. Section 2 introduces the concepts surrounding DUCT and the environment in which it has been implemented. Section 3 discusses DUCT's design, problems encountered during development, and possible enhancements. Two examples that highlight the novel features of DUCT and how it is useful when used within the relative debugging framework are illustrated in Section 4.

## 2. Background

DUCT locates the definitions of a requested variable by constructing use define (UD) chains [5,13]. A UD chain is a list of all definitions that reach a particular variable use within a program. By utilizing UD chains programmers can quickly and efficiently locate the origin of a faulty variable.

The majority of UD chain applications generate the exhaustive set of UD chains for a program by using global analysis techniques [5]. DUCT avoids the inherent expense of global analysis by adopting a demand driven approach [9,14]. Accordingly, DUCT only generates, as requested, UD chains for those variables that are of interest to the programmer. A demand driven approach allows DUCT to be efficiently used as time consuming, and possibly redundant, program analysis on the entire program is avoided. In addition, the generated information is cached so that it may be used during the construction of subsequent UD chains.

DUCT generates UD chains for programs that execute within the Microsoft .NET framework and clearly displays the result within the Microsoft Visual Studio environment [20]. A .NET program is contained within a portable executable (PE) file that contains compiler generated Microsoft Intermediate Language (MSIL) [19] and metadata [18] that conforms to the Common Type Specification (CTS) [17]. MSIL is a CPU independent instruction set that can be efficiently transformed into native code (by a just-in-time (JIT) compiler) at runtime. Metadata allows a .NET program to be self-describing by detailing the types, signatures, and other data that the .NET framework uses. DUCT has been integrated into Microsoft Visual Studio .NET using the Visual Studio Integration Program (VSIP) [21]. VSIP provides a framework in which the Visual Studio .NET architecture may be extended. The framework allows DUCT to be hosted by Microsoft Visual Studio .NET and utilize the IDE and services offered by the environment.

Most program analysis and comprehension tools are developed for use with one particular programming language [22,4,8]. DUCT constructs UD chains by performing the required analysis directly on the intermediate code and, therefore, avoids the tight coupling with the high level programming language. Accordingly, DUCT is not bound to one particular programming language, but rather, functions with any programming language that targets the intermediate language.

Recent investigation on static analysis of intermediate languages has focused on traditional global analysis [16,25,26]. However, performing program analysis on intermediate languages for interactive purposes has received little consideration. The work in [15] details the construction of control flow graphs, symbol table information, and UD chain data from the Java bytecode but, unlike DUCT, the technique is not interprocedural or demand driven.

## 3. DUCT

To locate the definition(s) reaching a variable the user must select the variable in the source program and inform DUCT to construct the UD chain. Currently, the user may select scalar variables, objects, object member fields, and arrays in the local scope for UD chain construction by DUCT. The reaching definition(s) located by DUCT are clearly highlighted in the source window and listed in an output pane that allows the user to easily navigation through the definitions. For example, Figure 1 illustrates the UD chain constructed by DUCT for the variable 'x' on line 22 in a simple C# program. Each definition that reaches the use of 'x' on line 22 is highlighted in the source code window and detailed in the output window at the bottom of the IDE. The user may navigate to a particular definition by clicking on the definition in the output window. This is particularly useful when a definition is not displayed on the screen or is located in another document.

To date, DUCT has been successfully tested with a number of different programming languages, in particular VB.NET, C#, C++ and GPCP [11].

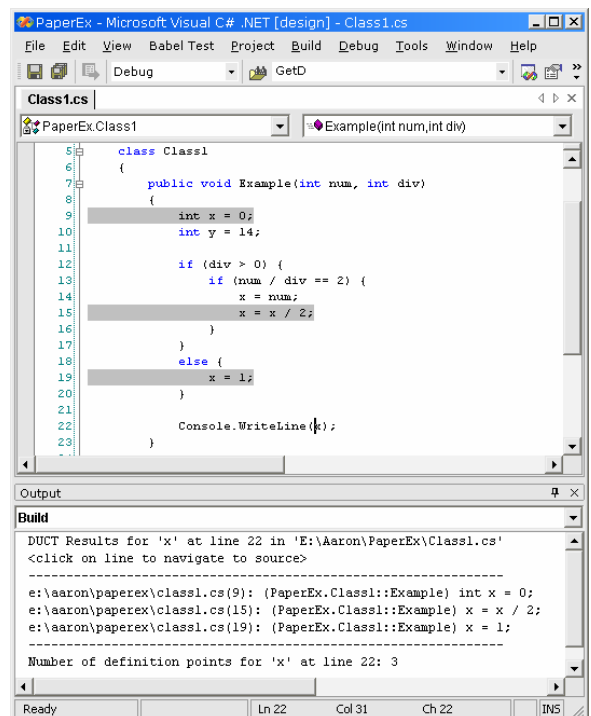

**Figure 1**

As discussed, the requested UD chain contains the definitions that reach the selected variable. As DUCT processes the intermediate language, to remain language independent, it cannot rely on the high level language grammar nor can it use the data structures, such as abstract syntax trees, symbol tables or control flow graphs, normally built and maintained by the compiler. DUCT therefore must build the data structures required to construct UD chains by processing the MSIL and metadata emitted by the compiler.

To construct the required information, DUCT utilizes the metadata and symbolic debug information when constructing UD chains. The metadata is used to obtain information about types that the code defines and the types that it references externally. The symbolic debug information, stored in the symbol store, contains the information that allows DUCT to interact with the user in relation to the high level source program. The data structures required by DUCT to construct UD chains are summarized in Table 1.

| Data Structure | Purpose |
| --- | --- |
| Control flow graph | A CFG provides the information that makes it possible to locate the last definition from each control path that leads to the selected variable. The CFG is constructed using conventional techniques [5]. |
| Call graph | DUCT builds a call graph [12] to allow UD chains to be constructed across method boundaries. A call graph represents the possible transfer of control between methods. |
| Class hierarchy | A class hierarchy [6,7] is constructed to allow DUCT to locate all possible definitions when a virtual call is encountered. That is, during UD chain construction each overriding method in each subtype must be scanned for possible variable definitions. |

**Table 1**

Once the required data structures have been constructed, DUCT can locate the variable definitions by identifying MSIL store instructions that update the memory allocated for the selected variable. The MSIL store instructions that DUCT must identify are detailed in table 2.

| Instruction | Description |
| --- | --- |
| starg.<length> *num* | Stores a value to the argument numbered *num*. The value is retrieved from the stack. |
| stelem.<type> | Stores a value in an element of an array. The array, element index, and value are retrieved from the stack. |
| stind.<type> | Stores a value of type <type> into memory. The memory address and value are retrieved from the stack. |
| stfld *field* | Stores a value into the field *field* of an object. The object and value are retrieved from the stack. |
| stloc.<index> | Stores a value into local variable *index*. The value is retrieved from the stack. |

**Table 2**

When a store instruction is located the target variable must be determined. All store instructions, except for indirect stores, implicitly encode the target variable and, hence, can be extracted directly from the instruction. In contrast, indirect store instructions receive the target address of the variable from the stack. Therefore, to correctly locate the load instruction that pushes the address of the target variable onto the stack DUCT must perform symbolic execution of the abstract stack machine.

To locate the definitions that reach the selected variable the CFG must be traversed back from the basic block that contains the selected variable. If this block contains a definition for the selected variable (prior to the use) it is the only definition that may reach the selected use (due to the properties of a basic block [5]). If the start block does not contain a definition for the selected variable it is possible that more than one definition may reach the use (along different control paths). Therefore, each control path that leads to the starting block needs to be scanned (backwards) for the last definition. This is achieved by (recursively) scanning the predecessors of the current block.

If a call instruction is encountered (along a path) during UD chain construction and the variable for which the UD chain is being constructed is an object, or passed by reference to the callee, then UD chain construction must continue at the called method. In this case, details of the called method are extracted (as outlined in step 3) and chain construction continues from the last block in the constructed CFG. The control paths in the called method are searched (backwards) for definitions of the corresponding formal parameter.

If each control path in the called method contains a definition for the corresponding formal parameter, then UD chain construction along the current path at the call-site can cease. However, if not every control path through the called method provides a definition for the corresponding formal parameter, UD chain construction must continue along the current control path at the call-site (from the instruction prior to the call).

In addition, if the call is to a virtual function then overriding methods in each subclass need to be proc-essed in the same manner. The class hierarchy constructed in step 3b is used to determine the overriding methods in each subclass.

If the beginning of the method has been reached (along a control path) during chain construction of a parameter then not all control paths within the method define a value for the incoming parameter. In this case, UD chain construction must continue at each callsite, defined in the constructed call graph, of the current method.

While the demand driven approach provides quick response times an initial, one-off, analysis of the entire program is required to construct the call graph. Specifically, an initial sweep of the program is required to locate the call instructions in order to build the call graph. The construction of the call graph is therefore linear to the size of the program. On the other hand, the class hierarchy can be quickly constructed from the information contained within the metadata and the CFG is only constructed a method is scanned for potential definitions.

## 4. Example

The benefits of using DUCT were highlighted when used to test and debug the *Earth* program [27] as it was upgraded from VB to VB.NET. The *Earth* program is a free program that uses the VSOP87 planetary theory to compute the heliocentric ecliptic longitude, latitude, and distance to the sun of the planet Earth over a period of several thousand years.

After running the original and upgraded versions with identical inputs we noted, as illustrated in Figure 2, that the resulting output was different. We proceeded to debug the error by locating the code that output the erroneous result (shown in Figure 3). This code indicated that the incorrect result was being produced by the value of variable *Q*. The value of *Q* is assigned its value by the function call to *EARTH_LBR_FOR* on the previous line. We placed an assertion on the input parameter, *Q*, to determine if the input parameter contained the same value in both versions.

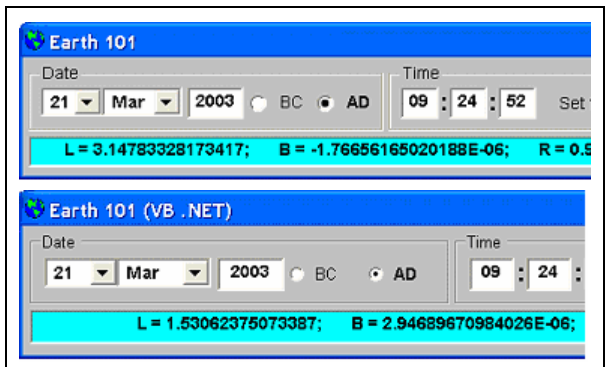

**Figure 2**

```
Private Sub ComputeButton_Click(
  ByVal eventSender As System.Object, _
  ByVal eventArgs As System.EventArgs) _
    Handles ComputeButton.Click

  Dim Q As Object

  JDE_FOR(INTERFACE_DATE, INTERFACE_TIME, Q)
  Q = EARTH_LBR_FOR(Q)
  Text2.Text = Q
End Sub
```
**Figure 3**

Next, we use DUCT to locate the definition(s) that assign a value to input parameter *Q*. Using the result produced by DUCT, illustrated in Figure 4, we place assertions on the variables, *W* and *Q*, used in the right hand expression of the definition located in the *JDE_FOR* function (shown in Figure 5).

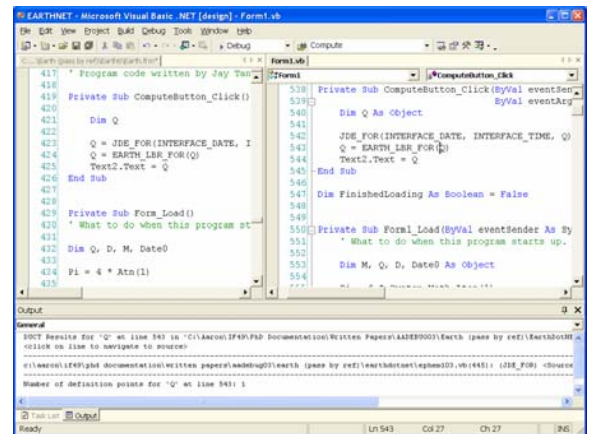

**Figure 4**

```
Public Function JDE_FOR(
  ByRef Date_String As Object, _
  ByRef Time_String As Object, _
  ByRef fracRes As Object)
  ' Returns the fraction of a day corresponding
to the given
  ' time argument in the standard
"HH:MM:SS.sss" format.

  Dim Q, W As Object

  JD_NUM_FOR(Trim(Date_String), W)
  Q = Trim(Time_String)
  Q = (Val(Left(Q, 2)) * 3600.0# + Val(Mid(Q,
4, 2)) * 60.0# + Val(Mid(Q, 7, 16))) / 86400.0#

  fracRes = W + Q

End Function
```
**Figure 5**

After setting these assertions we ran Guard to determine which variable contained an incorrect value. Guard identified, as shown in Figure 6, that the variable

*W* in the function *JDE_FOR* contained two different values in the two programs.

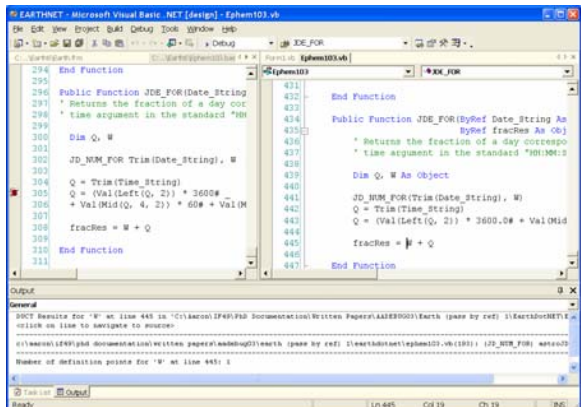

**Figure 6**

Using DUCT again we located the definition that assigns a value to *W* used in the statement *fracRes = W + Q* in the *JDE_FOR* function. The results are shown in Figure 7. Navigating to the definition located by DUCT, which resides in the *JD_NUM_FOR* function (shown in Figure 8), we continue the process and place assertions on the define points of variables MM, MMM, Pointer, Q and DD. This allows us to trace back through the expressions and determine that DD was correct but Len(DD) was incorrect. It transpired that in the Visual Basic 6 version of the code, the Len function only takes a string argument, and thus when it is passed a variant of type double this is first converted to a character string, and the length of that string is returns. However, in Visual Basic .NET, the Len function takes an Object as a parameter, and it returns the length of the object – in this case, a value of 8. The code can be corrected by explicitly converting DD to a string as shown in Figure 9.

**Figure 7**

```
Public Function JD_NUM_FOR(_
  ByRef DD_MMM_YYYY_BCAD As Object, _
  ByRef astroJDnum As Object)
     ...
  Date_String = Trim(UCase(DD_MMM_Yyyy_BCAD))
  Q = ""
  For Pointer = 1 To Len(Date_String)
    Q1 = Mid(Date_String, Pointer, 1)
    If Q1 <> " " Then Q = Q & Q1
  Next Pointer
```

```
  Date_String = Q
  DD = Val(Q)
  Pointer = InStr(1, Q, DD) + Len(DD)
  ...
  MMM = Mid(Q, Pointer, 3): Pointer = Pointer + 3
  MM = Int(1 + ((InStr(1, "JANFEBMARAPRMAYJUN-JULAUGSEPOCTNOVDEC", MMM) - 1) / 3))
  ...
  JD = DD + Int(367 * (MM + (Q * 12) - 2) / 12) + Int(1461 * (YYYY + 4800 - Q) / 4) - 32113
  ...
  astroJDnum = JD - 0.5

End Function
```

**Figure 8**

```
Pointer = InStr(1, Q, DD) + Len(CType(DD, Sys-
tem.String))
```

**Figure 9**

In this example, DUCT was instrumental in following the chain of errors back to the source. It allowed us to navigate the use-define chain quickly, and thus helped us to determine where to place assertions. *In all, it only took 4 iterations to locate the error, and a total of 11 assertions were required.*

## 5. Conclusion and Future Work

Relative debugging has been proven to be a powerful paradigm for testing and debugging programs that have undergone software evolution. Relative debugging currently requires the user to identify the data structures and locations within the two programs to define assertion commands. This process can be time consuming and error prone without a detailed knowledge of the programs under consideration. This paper has introduced a tool, DUCT, that allows a user to navigate the data flow of suspect variables throughout the programs. Such information provides valuable assistance when formulating assertion commands.

While DUCT is a valuable tool that assists the user during debugging it also exhibits a number of novel features. DUCT performs efficient demand driven program analysis on the intermediate language. This approach allows DUCT to be used with any high level programming language that targets the intermediate language.

Several limitations and known problems are the subject of ongoing development. Future work will include alias and array analysis. Traditionally, such analysis is global. Investigation will be required to determine an approach that lends itself to the demand driven approach and does not impose severe response times. Safety considerations dictate that any solution is conservative so that no definition point is ever missed.

DUCT does not currently handle delegates (function pointers). Initial investigation suggests that the most efficient way to process delegates is to locate the instantiations of every delegate during the initial program sweep that is required to construct the call graph and class hierarchy. Although this approach produces conservative UD chains, we believe the response time will still be adequate.

Instance and virtual methods are passed the instance on which the invocation operates. This parameter is usually referred to as the 'this' parameter in the language community but has various names in high level languages (i.e., 'this' in C#, 'self' in VB, etc). The metadata or symbol store do not contain the high level name that the 'this' parameter is referenced by, preventing DUCT from constructing UD chains for the 'this' object.

DUCT does not construct UD chains for global variables. Difficulties arise, in a multi-threaded program, because the definition(s) for a global variable may reside on different control paths than the one being considered. Investigation is required to determine the best approach to handle global variables.

We are also currently investigating the use of DUCT to automatically generate the assertions between two programs. Minimizing the user's involvement would reduce the cost of maintaining, enhancing, and porting software and have significant impact on current practices in software development.

## Acknowledgements

This work has been supported by a grant provided by the Australian Research Council under the Large Grants Scheme. We wish to acknowledge the support of Microsoft Corporation who provided access to the Visual Studio Integration Program (VSIP) whilst it was still in Beta form. Particular thanks go to Dan Fay, Todd Needham and Frank Goscinski of Microsoft.